\documentclass[sigconf]{acmart}

\usepackage{booktabs} % For formal tables

\usepackage{hyperref}
\makeatletter
\def\UrlAlphabet{%
	\do\a\do\b\do\c\do\d\do\e\do\f\do\g\do\h\do\i\do\j%
	\do\k\do\l\do\m\do\n\do\o\do\p\do\q\do\r\do\s\do\t%
	\do\u\do\v\do\w\do\x\do\y\do\z\do\A\do\B\do\C\do\D%
	\do\E\do\F\do\G\do\H\do\I\do\J\do\K\do\L\do\M\do\N%
	\do\O\do\P\do\Q\do\R\do\S\do\T\do\U\do\V\do\W\do\X%
	\do\Y\do\Z}
\def\UrlDigits{\do\1\do\2\do\3\do\4\do\5\do\6\do\7\do\8\do\9\do\0}
\g@addto@macro{\UrlBreaks}{\UrlOrds}
\g@addto@macro{\UrlBreaks}{\UrlAlphabet}
\g@addto@macro{\UrlBreaks}{\UrlDigits}
\makeatother

\renewcommand\footnotetextcopyrightpermission[1]{} % removes footnote with conference information in first column

\begin{document}
	
%	\copyrightyear{2018} 
%	\acmYear{2018} 
%	\setcopyright{acmcopyright}
%	\acmConference[CIKM '18]{2018 ACM Conference on Information and Knowledge Management}{October 22--26, 2018}{Torino, Italy}
%	\acmBooktitle{2018 ACM Conference on Information and Knowledge Management (CIKM'18), October 22--26, 2018, Torino, Italy}
%	\acmPrice{15.00}
%	\acmDOI{10.1145/XXXXXX.XXXXXX}
%	\acmISBN{978-1-4503-6014-2/18/10} 
%	% Authors, replace the red X's with your assigned DOI string during the rightsreview eform process.
	
	\fancyhead{}

\title{False News Detection on Social Media}

%\author{Juan Cao}
%\email{caojuan@ict.ac.cn}

\author{Juan Cao$^{1,2}$, Qiang Sheng$^{1,2}$, Peng Qi$^{1,2}$,  Lei Zhong$^{1,2}$, Yanyan Wang$^{1,2}$ and Xueyao Zhang$^{1,2}$}
\affiliation{%
	\institution{
		$^{1}$Key Laboratory of Intelligent Information Processing \& 
		Center for Advanced Computing Research,\\
		Institute of Computing Technology, CAS, Beijing, China\\
		$^{2}$University of Chinese Academy of Sciences, Beijing, China}
}
\email{{caojuan,shengqiang,qipeng,zhonglei,wangyanyan,zhangxueyao}@ict.ac.cn}

\begin{abstract}
% Background
Social media has become an important information platform where people consume and share news. However, it has also enabled the wide dissemination of false news, i.e., news posts published on social media that are verifiably false, causing significant negative effects on society. 
To help prevent further propagation of false news on social media, we set up this competition to motivate the development of automated real-time false news detection approaches.
Specifically, this competition includes three subtasks: false-news text detection, false-news image detection, and false-news multi-modal detection, which aims to motivate participants to further explore the efficiency of multiple modalities in detecting false news and effective fusion approaches of multi-modal contents.
To better support this competition, we also construct and release a multi-modal data repository about \textit{\underline{F}alse \underline{Ne}ws on \underline{W}eibo \underline{S}ocial platform(MCG-FNeWS}) to help evaluate the performance of different approaches from participants.

\end{abstract}

%\begin{CCSXML}
%<ccs2012>
%% <concept>
%%  <concept_id>10010520.10010553.10010562</concept_id>
%%  <concept_desc>Computer systems organization~Embedded systems</concept_desc>
%%  <concept_significance>500</concept_significance>
%% </concept>
%% <concept>
%%  <concept_id>10010520.10010575.10010755</concept_id>
%%  <concept_desc>Computer systems organization~Redundancy</concept_desc>
%%  <concept_significance>300</concept_significance>
%% </concept>
%</ccs2012>
%\end{CCSXML}
%
%\ccsdesc[500]{Computer systems organization~Embedded systems}

%\keywords{datasets, neural networks, gaze detection, text tagging}

\maketitle

\section{Introduction}
% background of false news: social media->definition->harmness
Social media, such as Twitter\footnote{https://twitter.com/} or Chinese Sina Weibo\footnote{https://weibo.com/}, has become an important information platform where people acquire the latest news and express their opinions freely \cite{socialmediause}, \cite{socialmediausechinese}. 
However, the convenience and openness of social media have also promoted the proliferation of false news, i.e., news posts published on social media that are verifiably false, which not only disturbed the cyberspace order but also caused many detrimental effects on real-world events.
For example, in India, dozens of innocent people were beaten to death by locals because of the false news about child trafficking that was widely spread on social media \cite{indiawhatsapp}.
Thus, false news detection is a critical issue that needs to be addressed.

% task: real-time false news detection
% challenge: early detection
Some existing researches utilize the information generated in the news proliferation process, such as reviews, retweets and other relevant posts, to help detect false news \cite{ma2016ijcai}, \cite{cikm2018rumor}, \cite{shu2019defend}, \cite{guoemotion}, 
but these contents can become available only after the news has been propagated on social networks for a while.
%information usually are generated 
%real news diffused significantly farther, faster, deeper, and more broadly than truth. In detail, 
However, according to statistics, false news spreads very quickly on social media, even six times faster than real news \cite{science2018spread}.
This further indicates that false news may have already been widely spread and caused many negative effects when enough relevant posts are generated.
%Along this line, one interesting problem is to perform early fake news detection, which aims to give early alerts of fake news during the dissemination process.
%For example, this approach could look at only social media posts within some time delay of the original post as sources for news verification [37]. 
Therefore, to help prevent further propagation of false news on social media, we set up this competition to motivate the development of automated real-time false news detection approaches.

% use raw news content to detect
To effectively detect the false news from the news feed on social media in real time, the information we are looking at will mostly be the raw news content, which mainly includes text, images or videos, and publisher profile.
% Solution
% 1. multimodality
% text related work
Traditional false news detection researches based on news content usually focus on the textual content \cite{textcastillo2011information}, \cite{ma2016ijcai}, \cite{ma2019wwwgan}, from where they exploit some linguistic features to capture the differences of writing styles between false and real news.
% mutlimodality
%In addition to textual content, visual cues have been shown to be an important manipulator for false news propaganda[].
With the evolution of self-media news from text-based posts to multimedia posts with images or videos, false news usually utilize misrepresented or even tampered images to attract and mislead readers for rapid dissemination, which leads researchers to pay more attention to the visual content of false news \cite{jin2015TMM}, \cite{icdm2019}.
Considering that multiple modalities could provide cues for distinguishing false news, some works propose novel models to fuse features from different modalities to solve the challenging false news detection problem \cite{jin2017multimodal}, \cite{eannwang2018}, \cite{mvae}.
% three subtask
For this reason, we set up this competition to encourage participants to fully utilize the raw news content for false news detection, which consists of three subtasks: (a) false-news text detection, (b) false-news image detection, and (c) false-news multi-modal detection.
% existing dataset
% multimodal dataset: mediaeval, mm17
Existing datasets about false news detection usually lack corresponding visual content \cite{datasettwitter15}, \cite{ma2016ijcai}, \cite{fakenewsnet}, and the scale of multi-modal datasets in this field are limited \cite{mediaeval15}, \cite{mediaeval16}, \cite{jin2017multimodal}.
% our dataset
Therefore, to better support this competition, we construct and publicize a multi-modal data repository about \textit{\underline{F}alse \underline{Ne}ws on \underline{W}eibo \underline{S}ocial platform(MCG-FNeWS}), which is the largest multi-modal false news detection dataset, to help evaluate the performance of different approaches from participants.
% 2. knowledge-based
Besides, external knowledge is also helpful for determining the truthfulness of a particular claim in a real-time \cite{surveyKai2017Fake}.
% resources
Thus, we also provide some resources which contain a large number of refutations about existing false news.
We encourage participants to utilize the given external knowledge to help detect false news. 

%添加不同子任务的相关研究背景 

\section{Task Overview}
%1. utilize raw news post information to make real-time detection, including textual, visual content
The problem addressed in this competition is how to utilize the raw news content, mainly including the textual and visual content and publisher profile, to verify whether the given post is false or real in real-time.
%% 加example
%2. different modalities play different roles, respectively establish subtask A and subtask B to explore the efficiency of different modalities in detecting
It has been proved that textual and visual content play important roles in detecting false news, thus we establish subtask A and subtask B to explore the efficiency of textual and visual modalities in detecting false news, respectively. 
%3. different modalities hubu zengqiang, subtask C explore how to effective fuse different modal 
Different modalities can not only mutually support but also be supplementary \cite{multimodalsurvey}, but how to effectively process and relate information from different modalities is still a challenging problem. 
Subtask C aims to effectively fuse the information of different modalities to detect false news.
%4. in all above subtasks, we encourage participants to fully utilize the external knowledge to help detection 
In all the above subtasks, we encourage participants to fully utilize the external knowledge that we have given to help detect false news.

% 加跟MediaEval的对比

\subsection{Subtask A - False-news text detection}
%1. text traditional key factor, deeply patterns have not been explored
Text is a major component of a news event, which is widely utilized by existing researches to verify the given news post is real or false. 
Many linguistic-based features have been widely studied to help to detect false news, but the underlying characteristics of false news have not been fully understood.  
%2. subtask A aim: explore the efficiency on detection
Therefore, the aim of subtask A is to further explore the efficiency of text content in detecting false news.
%4. success can assist on subtask C
Success on this subtask will support the success of subtask C by providing effective features.
%3. definition false-news text
The definition of subtask A is the following:
%"Given a social media post, comprising a text component solely, the subtask requires participants to return a decision on whether the given post is real-news or false-news."
"Given a set of news posts $\mathcal{X}=\left\{x_{1}, x_{2}, \dots, x_{m}\right\}$ and labels $\mathcal{Y}=\left\{y_{1}, y_{2}, \dots, y_{m}\right\}$, the subtask requires participants to learn a classifier $f$ that can utilize the corresponding text to classify whether a given post is false news ($y_{t}=1$) or real news ($y_{t}=0$) , i.e., $\hat{y_t} = f(x_t)$. "
Accordingly, we define \textbf{\textit{false-news text}} as text in false news, and \textit{real-news text} as text in real news. 
In practice, 
participants receive a list of text and are required to automatically predict, for each text, whether it is a false-news text or a real-news text.
%5. example
%Figure * gives examples of the material.

\subsection{Subtask B - False-news image detection}
% 1. visual cues important manpulator for fake news, limited research
Visual cues have been shown to be an important manipulator for false news detection\cite{2013fakeimages}, \cite{mediaverify}.
However, very limited research has been done to exploit effective visual features, including traditional local and global features \cite{imagefea2013review} and newly emerging deep network-based features \cite{icdm2019}, for the false news detection problem.
% 2. the aims, support task C
Subtask B encourages the participants to put more attention on the visual content (images) to detect false news.
Similarly, success of this subtask also promotes the success of subtask C.
% 3. definition
The definition of subtask B is the following:
Given a set of news posts $\mathcal{X}=\left\{x_{1}, x_{2}, \dots, x_{m}\right\}$, corresponding images $\mathcal{I}=\left\{i_{1}, i_{2}, \dots, i_{m}\right\}$, and labels $\mathcal{Y}=\left\{y_{1}, y_{2}, \dots, y_{m}\right\}$, learn a classifier $f$ that can utilize the corresponding image to classify whether a given post is false news ($y_{t}=1$) or real news ($y_{t}=0$) , i.e., $\hat{y_t} = f(i_t)$. 
Accordingly, we define \textbf{\textit{false-news image}} as attached image in false news, and \textit{real-news image} as attached image in real news. 
In practice, 
participants receive a list of images and are required to automatically predict, for each image, whether it is a false-news image or a real-news image.
Note that this subtask is different from tampered image detection because the tampered image is only a typical category of false-news image \cite{icdm2019}. 
% for example

% 跟fake images的区别

\subsection{Subtask C - False-news multi-modal detection}
This subtask aims at utilizing information from different modalities to effectively detect false news.
Although there are already some studies focusing on fusing multi-modal information for
false news detection, it is still a challenging problem which needs further investigation. For example, we can use the semantic alignment between image and text to explore the role
of different modalities in false news detection, or utilize the technique of co-learning to tackle the problem of missing data.
% definition
The definition of subtask C is the following:
Given a set of news posts $\mathcal{X}=\left\{x_{1}, x_{2}, \dots, x_{m}\right\}$, corresponding images $\mathcal{I}=\left\{i_{1}, i_{2}, \dots, i_{m}\right\}$, publisher profile $\mathcal{U}=\left\{u_{1}, u_{2}, \dots, u_{m}\right\}$, and labels $\mathcal{Y}=\left\{y_{1}, y_{2}, \dots, y_{m}\right\}$, learn a classifier $f$ that can utilize the corresponding text, image and publisher profile to classify whether a given post is false news ($y_{t}=1$) or real news ($y_{t}=0$) , i.e., $\hat{y_t} = f(x_t, i_t, u_t)$. 
% topic information
Moreover, we refer to existing category lists from well-known debunking websites and finally summarize the following nine overarching topics: Society \& Life, Disasters \& Accidents, Health \& Medicine, Education \& Examinations, Science \& Technology, Finance \& Business, Culture \& Sports \& Entertainment, Politics and Military. 
For each post in the dataset, we also provide a topic tag which is manually labeled by its key objects of interest.
In practice, 
participants receive a list of posts which include a text component, an associated images list, a user profile, and a topic tag, and are required to automatically predict, for each post, whether it is a false-news post or a real-news post.

In all cases, the competition asks participants to optionally return an explanation (which can be a text string, or indexes pointing to the given knowledge) that supports the verification decision. The explanation is not used for quantitative evaluation, but rather for gaining qualitative insights into the results.

%\section{Verification Corpus}
\section{Data \& Resources}
% overview: label, amount
\textbf{Training dataset: }This is provided with ground truth and is used by participants to develop their approaches. It contains 38,471 news posts with 34,096 corresponding images, comprising 19,285 false-news posts with corresponding 13,635 false-news images, and 19,186 real-news posts with corresponding 20,461 real-news images. 
\\
\textbf{Validation dataset: }This is provided with ground truth and is used by participants to evaluate their approaches. It contains 4,000 news posts with 3,837 corresponding images, comprising 2,000 false-news posts with corresponding 1,760 false-news images, and 2,000 real-news posts with corresponding 2,077 real-news images. 
\\
\textbf{Testing dataset: }This is provided without ground truth and is used by organizers to compare the performance of participants' approaches. It contains 3,902 news posts with 3,957 corresponding images.
%, comprising 1902 false-news posts with corresponding 1742 false-news images, and 20t00 true-news posts with corresponding 2215 true-news images. 

In all datasets, the text of false news and real news are used to develop subtask A, images are used to develop subtask B, and all given data are for subtask C.

% source, crawel strategy, preprocess
The data for all datasets are publicly available\footnote{https://www.biendata.com/competition/falsenews/data/}. 
%fake: piyaopingtai, time span, remove text empty
The false-news posts are crawled from May 2012 to November 2018 and verified by the official Weibo Community Management Center\footnote{https://service.account.weibo.com/}, which usually serves as a reputable source to collect false-news posts on Weibo platform in literature \cite{ma2016ijcai}, \cite{jin2017multimodal}, \cite{liu2017rumors}, \cite{zhao2018fake}. 
% text 为空还有图片呀，use profile无法抓取
%truth: same time span
% datamining->fake news pattern,search, keywords search, remove duplicate, rengong label, select same amount, event-adversial
The real-news posts are collected during the same period as false news from Weibo. 
To explore the underlying characteristics of false-news posts in addition to superficial linguistics features, we crawl some real-news posts which have the similar linguistic style with false-news posts as negative samples. 
Specifically, 
following the method in \cite{jin2016arxivimage}, we discover false-news linguistics patterns like "is it real/false?" in false-news posts via text mining, and then crawl a large set of matched posts from the live stream of Weibo. 
For each post, we extract the keywords as the seed to crawl corresponding posts.
After removing the duplicated posts, we obtain a candidate set of real-news posts, 
which are further manually verified by cross-checking online sources(articles and blogs), producing a real-news set.
Finally, we sample the real-news posts to keep the balance of false-news and real-news posts.
To alleviate the impact of events \cite{eannwang2018}, we select real-news posts that belong to the same or similar events with false-news posts.   
% preprocess
In the preprocessing stage, we manually remove some meaningless statistical clues from the text.  

%1. user profile 和category写了吗？
%2. image是在筛选后的text里做的去重吗？

% Format/example

% Statistics 可写可不写
We also provide a debunking repository which contains 37,877 refutations about existing false news. 
We crawl these refutations from multiple reputable debunking Weibo accounts and web articles.
These refutations are crawled from September 2012 to August 2019.
We encourage participants to utilize these refutations to help the detection of false news, but we do not promise that all false news in the competition dataset has corresponding refutations in this debunking repository. 
% 不保证包含fake news所包含的
% 微博辟谣账号， 政务机构，媒体，第三方的澄清账号
% 最早月份

\section{Evaluation}
Overall, all the above subtasks are interested in the accuracy with which an automatic method can distinguish between false news and real news. Hence, given the testing set of labeled instances and a set of predicted labels (included in the submitted runs) for these instances, the classic measures (i.e., Precision P, Recall R, and F1-score) are used to quantify the classification performance, where the target class is the class of false news. Since the two classes (false news/real news) are represented in a relatively balanced way in the testing set, these measures are good proxies of the classifier accuracy.

\section{Baselines}
In this section, we provide some baselines of the three subtasks for reference, which are shown in Table 1.
For each subtask, we deploy some basic and state-of-the-art baselines on given datasets. 
Note that we doesn't focus on searching the best hyper-parameters of these model, thus the given baselines are not the best results of corresponding models. 
\begin{itemize}
	\item \textbf{Subtask A: } For subtask A, we introduce four basic models including LSTM \cite{lstm}, GRU \cite{gru}, TextCNN \cite{textcnn} and  Bert \cite{bert}, which are widely used in many NLP applications.
	% Bert??????
	In detail, we adopt the implementation of Bert in \cite{bertimplementation}. 
	According to Table 1, Bert is slightly better than other models in accuracy. 
	\item \textbf{Subtask B: } VGG\cite{vgg} is widely used as a feature extractor in existing studies about multi-modal fake news detection\cite{jin2017multimodal}, \cite{eannwang2018}, \cite{mvae}, thus we implement pre-trained and fine-tuned VGG19 as baselines of subtask B.
	Also, we implement the state-of-the-art method utilizing visual content to detect false news MVNN \cite{icdm2019}, which is much better than other baselines in subtask B.
	\item \textbf{Subtask C: } 
	For subtask C, we introduce three baselines including early and late fusion and attRNN\cite{jin2017multimodal} to fuse the information of text, image and user modality.
	Specifically,  we use TextCNN and pre-trained VGG19 to extract the abstract representations of text and image respectively. 
	Early fusion integrates features from different modalities by simply concatenating their representations, while late fusion performs integration after each of the modalities has made a classification decision. 
	More intuitively, attRNN proposes a neuron-level attention mechanism to fuse multi-modal content. 
	According to Table 1, early fusion outperforms other baselines for Subtask C.
\end{itemize}

\begin{table}
	\caption{Baselines for Three Subtasks}
	\label{tab:freq}
	\begin{tabular}{lcccc}
		\toprule
		Method&Accuracy&Precision&Recall&F1\\
		\midrule
		LSTM & 0.864 & 0.891 & \textbf{0.829} & \textbf{0.859} \\
%		LSTM + Attention & 0.815 & 0.961 & 0.647 & 0.773 \\
		GRU & 0.857 & 0.911 & 0.784 & 0.843 \\
%		GRU + Attention & 0.832 & 0.950 & 0.691 & 0.800 \\
		TextCNN & 0.851 &\textbf{ 0.953} & 0.732 & 0.828 \\
		Bert & \textbf{0.867} & 0.916 & 0.799 & 0.854 \\
		\midrule
		Pre-trained VGG19 & 0.728 & 0.729 & 0.622 & 0.671 \\
		Fine-tuned VGG19 & 0.759 & 0.791 & 0.607 & 0.687 \\
		MVNN & \textbf{0.805} & \textbf{0.804} & \textbf{0.743} & \textbf{0.772} \\
		\midrule
		Early Fusion & \textbf{0.876} & 0.916 & \textbf{0.837} & \textbf{0.875} \\
		Late Fusion & 0.846 & \textbf{0.935 }& 0.757 & 0.836 \\
		attRNN & 0.852 & 0.871& 0.820 & 0.845 \\
		\bottomrule
	\end{tabular}
\end{table}

\section{Conclusion}
With the popularity of multi-modal content in social media, incorporating the information of different modalities to detect false news is a critical task in the current media landscape. 
This competition about false news detection set up three subtasks to encourage participants to fully explore the efficiency of different modalities and effective fusion methods.
This competition also leaves behind a benchmark dataset of ten thousands of false news and real news, which will help beginners of this research domain to quickly get started and evaluate their systems.

\section{Acknowledgments}
This work was supported by the National Natural Science Foundation of China(U1703261).

\bibliographystyle{unsrt}
\bibliography{ref} 

\begin{thebibliography}{10}

\bibitem{socialmediause}
Timothy~I Murphy.
\newblock News use across social media platforms 2018.
\newblock
  \url{https://www.journalism.org/2018/09/10/news-use-across-social-media-platforms-2018/}.
\newblock Accessed September 10, 2018.

\bibitem{socialmediausechinese}
A research report about china internet news market 2016.
\newblock
  \url{http://www.cnnic.cn/hlwfzyj/hlwxzbg/mtbg/201701/t20170111_66401.htm}.
\newblock Accessed January 11, 2017.

\bibitem{indiawhatsapp}
Annie Gowen.
\newblock As mob lynchings fueled by whatsapp messages sweep india, authorities
  struggle to combat fake news.
\newblock
  \url{https://www.washingtonpost.com/world/asia_pacific/as-mob-lynchings-fueled-by-whatsapp-sweep-india-authorities-struggle-to-combat
  -fake-news/2018/07/02/683a1578-7bba-11e8-ac4e-421ef7165923_story.html?noredirect=on}.
\newblock Accessed July 2, 2018.

\bibitem{ma2016ijcai}
Jing Ma, Wei Gao, Prasenjit Mitra, Sejeong Kwon, Bernard~J Jansen, Kam-Fai
  Wong, and Meeyoung Cha.
\newblock Detecting rumors from microblogs with recurrent neural networks.
\newblock In {\em IJCAI}, pages 3818--3824, 2016.

\bibitem{cikm2018rumor}
Han Guo, Juan Cao, Yazi Zhang, Junbo Guo, and Jintao Li.
\newblock Rumor detection with hierarchical social attention network.
\newblock In {\em Proceedings of the 27th ACM International Conference on
  Information and Knowledge Management}, pages 943--951. ACM, 2018.

\bibitem{shu2019defend}
Kai Shu, Limeng Cui, Suhang Wang, Dongwon Lee, and Huan Liu.
\newblock defend: Explainable fake news detection.
\newblock 2019.

\bibitem{guoemotion}
Chuan Guo, Juan Cao, Xueyao Zhang, Kai Shu, and Miao Yu.
\newblock Exploiting emotions for fake news detection on social media, 2019.

\bibitem{science2018spread}
Soroush Vosoughi, Deb Roy, and Sinan Aral.
\newblock The spread of true and false news online.
\newblock {\em Science}, 359(6380):1146--1151, 2018.

\bibitem{textcastillo2011information}
Carlos Castillo, Marcelo Mendoza, and Barbara Poblete.
\newblock Information credibility on twitter.
\newblock In {\em Proceedings of the 20th international conference on World
  wide web}, pages 675--684. ACM, 2011.

\bibitem{ma2019wwwgan}
Jing Ma, Wei Gao, and Kam-Fai Wong.
\newblock Detect rumors on twitter by promoting information campaigns with
  generative adversarial learning.
\newblock 2019.

\bibitem{jin2015TMM}
Zhiwei Jin, Juan Cao, Yongdong Zhang, Jianshe Zhou, and Qi~Tian.
\newblock Novel visual and statistical image features for microblogs news
  verification.
\newblock {\em IEEE transactions on multimedia}, 19(3):598--608, 2017.

\bibitem{icdm2019}
Peng Qi, Juan Cao, Tianyun Yang, Junbo Guo, and Jintao Li.
\newblock Exploiting multi-domain visual information for fake news detection.
\newblock In {\em 19th IEEE International Conference on Data Mining}. IEEE,
  2019.

\bibitem{jin2017multimodal}
Zhiwei Jin, Juan Cao, Han Guo, Yongdong Zhang, and Jiebo Luo.
\newblock Multimodal fusion with recurrent neural networks for rumor detection
  on microblogs.
\newblock In {\em Proceedings of the 2017 ACM on Multimedia Conference}, pages
  795--816. ACM, 2017.

\bibitem{eannwang2018}
Yaqing Wang, Fenglong Ma, Zhiwei Jin, Ye~Yuan, Guangxu Xun, Kishlay Jha, Lu~Su,
  and Jing Gao.
\newblock Eann: Event adversarial neural networks for multi-modal fake news
  detection.
\newblock In {\em Proceedings of the 24th ACM SIGKDD International Conference
  on Knowledge Discovery \& Data Mining}, pages 849--857. ACM, 2018.

\bibitem{mvae}
Khattar Dhruv, Goud Jaipal~Singh, Gupta Manish, and Varma Vasudeva.
\newblock Mvae: Multimodal variational autoencoder for fake news detection.
\newblock In {\em Proceedings of the 2019 World Wide Web Conference}. ACM,
  2019.

\bibitem{datasettwitter15}
Xiaomo Liu, Armineh Nourbakhsh, Quanzhi Li, Rui Fang, and Sameena Shah.
\newblock Real-time rumor debunking on twitter.
\newblock In {\em Proceedings of the 24th ACM International on Conference on
  Information and Knowledge Management}, pages 1867--1870. ACM, 2015.

\bibitem{fakenewsnet}
Kai Shu, Deepak Mahudeswaran, Suhang Wang, Dongwon Lee, and Huan Liu.
\newblock Fakenewsnet: A data repository with news content, social context and
  dynamic information for studying fake news on social media.
\newblock {\em arXiv preprint arXiv:1809.01286}, 2018.

\bibitem{mediaeval15}
Christina Boididou, Katerina Andreadou, Symeon Papadopoulos, Duc-Tien
  Dang-Nguyen, Giulia Boato, Michael Riegler, Yiannis Kompatsiaris, et~al.
\newblock Verifying multimedia use at mediaeval 2015.
\newblock In {\em MediaEval}, 2015.

\bibitem{mediaeval16}
Christina Boididou, Symeon Papadopoulos, Duc-Tien Dang-Nguyen, Giulia Boato,
  Michael Riegler, Stuart~E. Middleton, Andreas Petlund, Yiannis Kompatsiaris,
  et~al.
\newblock Verifying multimedia use at mediaeval 2016.
\newblock In {\em MediaEval}, 2016.

\bibitem{surveyKai2017Fake}
Shu Kai, Suhang Wang, Amy Sliva, Jiliang Tang, and Huan Liu.
\newblock Fake news detection on social media: A data mining perspective.
\newblock {\em Acm Sigkdd Explorations Newsletter}, 19(1), 2017.

\bibitem{multimodalsurvey}
Tadas Baltru{\v{s}}aitis, Chaitanya Ahuja, and Louis-Philippe Morency.
\newblock Multimodal machine learning: A survey and taxonomy.
\newblock {\em IEEE Transactions on Pattern Analysis and Machine Intelligence},
  41(2):423--443, 2018.

\bibitem{2013fakeimages}
Aditi Gupta, Hemank Lamba, Ponnurangam Kumaraguru, and Anupam Joshi.
\newblock Faking sandy: characterizing and identifying fake images on twitter
  during hurricane sandy.
\newblock In {\em Proceedings of the 22nd international conference on World
  Wide Web}, pages 729--736. ACM, 2013.

\bibitem{mediaverify}
Petter~Bae Brandtzaeg, Marika L{\"u}ders, Jochen Spangenberg, Linda
  Rath-Wiggins, and Asbj{\o}rn F{\o}lstad.
\newblock Emerging journalistic verification practices concerning social media.
\newblock {\em Journalism Practice}, 10(3):323--342, 2016.

\bibitem{imagefea2013review}
Dong ping Tian et~al.
\newblock A review on image feature extraction and representation techniques.
\newblock {\em International Journal of Multimedia and Ubiquitous Engineering},
  8(4):385--396, 2013.

\bibitem{liu2017rumors}
Yahui Liu, Xiaolong Jin, Huawei Shen, and Xueqi Cheng.
\newblock Do rumors diffuse differently from non-rumors? a systematically
  empirical analysis in sina weibo for rumor identification.
\newblock In {\em Pacific-Asia Conference on Knowledge Discovery and Data
  Mining}, pages 407--420. Springer, 2017.

\bibitem{zhao2018fake}
Zilong Zhao, Jichang Zhao, Yukie Sano, Orr Levy, Hideki Takayasu, Misako
  Takayasu, Daqing Li, and Shlomo Havlin.
\newblock Fake news propagate differently from real news even at early stages
  of spreading.
\newblock {\em arXiv preprint arXiv:1803.03443}, 2018.

\bibitem{jin2016arxivimage}
Zhiwei Jin, Juan Cao, Jiebo Luo, and Yongdong Zhang.
\newblock Image credibility analysis with effective domain transferred deep
  networks.
\newblock {\em arXiv preprint arXiv:1611.05328}, 2016.

\bibitem{lstm}
Sepp Hochreiter and J{\"u}rgen Schmidhuber.
\newblock Long short-term memory.
\newblock {\em Neural computation}, 9(8):1735--1780, 1997.

\bibitem{gru}
Junyoung Chung, Caglar Gulcehre, KyungHyun Cho, and Yoshua Bengio.
\newblock Empirical evaluation of gated recurrent neural networks on sequence
  modeling.
\newblock {\em arXiv preprint arXiv:1412.3555}, 2014.

\bibitem{textcnn}
Yoon Kim.
\newblock Convolutional neural networks for sentence classification.
\newblock {\em arXiv preprint arXiv:1408.5882}, 2014.

\bibitem{bert}
Jacob Devlin, Ming-Wei Chang, Kenton Lee, and Kristina Toutanova.
\newblock Bert: Pre-training of deep bidirectional transformers for language
  understanding.
\newblock {\em arXiv preprint arXiv:1810.04805}, 2018.

\bibitem{bertimplementation}
Implementation of bert that could load official pre-trained models for feature
  extraction and prediction.
\newblock \url{https://github.com/CyberZHG/keras-bert}.

\bibitem{vgg}
Karen Simonyan and Andrew Zisserman.
\newblock Very deep convolutional networks for large-scale image recognition.
\newblock {\em arXiv preprint arXiv:1409.1556}, 2014.

\end{thebibliography}

\end{document}